\documentclass[aps,prb,twocolumn,showpacs,amsmath]{revtex4}
\usepackage{amsmath}
\usepackage{latexsym}
\usepackage{amssymb}
\usepackage{graphicx}
\usepackage{bm}

\newcommand{\pdag}{{\phantom{\dagger}}}
\newcommand{\bq}{\begin{equation}}
\newcommand{\eq}{\end{equation}}
\newcommand{\bn}{\begin{eqnarray}}
\newcommand{\en}{\end{eqnarray}}

\begin{document}

\title{Controlling inelastic cotunneling through an interacting quantum dot by a 
circularly-polarized field}

\author{Bing Dong} 
\affiliation{Department of Physics, Shanghai Jiaotong University,
1954 Huashan Road, Shanghai 200030, China}

\author{X. L. Lei} 
\affiliation{Department of Physics, Shanghai Jiaotong University,
1954 Huashan Road, Shanghai 200030, China}

\begin{abstract}

We study inelastic cotunneling through a strong Coulomb-blockaded quantum dot subject to a 
static magnetic field and a perpendicular circularly-polarized magnetic field using a quantum 
Langevin equation approach. 
Our calculation predicts an interesting controllable cotunneling current characteristic, 
splitting--zero-anomaly--splitting transition of the differential conductance with increasing 
the driving frequency, ascribing to the role of {\em photon-assisted spin-flip} cotunneling 
processes.       

\end{abstract}

\pacs{72.40.+w, 73.63.Kv, 72.10.Fk}

\today

\maketitle

Coherent control of spin dynamics in a quantum dot (QD) using circularly-polarized magnetic 
field (CPF)\cite{Rabi,Grifoni} in a solid-state environment has been a subject of increasing 
interest in recent years.\cite{Awschalom} This interest has led to a large amount of work on 
detection of the electron spin resonance (ESR) in a QD,\cite{ESR1,ESR2,Engel,Gywat} and 
initializing the state of electron spin via optical absorption of circularly polarized 
light\cite{CPLop,Shabaev,Pryor}. Recently, the single-electron ESR in a QD-lead system with 
sizable Zeeman splitting has been theoretically reported to carry pure spin flow, which can 
be used as a fundamental element and/or a spin source device in the spintronic 
circuit.\cite{Zhao,spinbattery} These papers have studied the transport through a QD in 
resonant tunneling regime by the nonequilibrium Green's function (NGF)\cite{Zhao} and the 
quantum rate equation approach,\cite{spinbattery} respectively. The effect of the s-d 
exchange interaction between the localized spin and conduction electrons on electron 
transmission probability in linear transport regime has been exploited for a mesoscopic ring 
embedded with an magnetic impurity by the NGF.\cite{Aldea}  
In this paper, we focus our studies, for the first time, on the influence of a driving CPF 
upon cotunneling (non-resonant coherent tunneling) through a strongly interacting QD in the 
weak-tunneling limit,\cite{Matveev,cotunnelingExp} predicting that the cotunneling current 
can be easily controlled by tuning the driving frequency.  

Cotunneling through a single-level $\epsilon_d$ interacting QD subject to an ambient constant 
magnetic field ${\bf B}_0$ along the $z$-axis and a driving external (magnetic) CPF with 
frequency $\omega_c$ whose direction rotates in the plane perpendicular to the $z$-axis, 
${\bf B}_1(t)=B_1 (\cos \omega_c t, \sin \omega_c t,0)$, can be described by the Hamiltonian 
$H=H_{B}+H_{0}+H_{\rm I}$:\cite{Matveev}
\begin{subequations}
\label{hamiltonian}
\bn
H_{B}&=& \sum_{\eta \mathbf{k}\sigma }\varepsilon _{\eta \mathbf{k}}c_{\eta
\mathbf{k}\sigma }^{\dag }c_{\eta \mathbf{k}\sigma }^{{\phantom{\dagger}}}, \\
H_{0}&=&  - \Delta_0 S^{z} - \frac{1}{2} \Delta ( e^{i\theta} S^+ + e^{-i\theta} 
S^-),\label{QD} \\ 
H_{\mathrm{I}}&=& \sum_{\eta \eta ^{\prime },\mathbf{k}
\mathbf{k}^{\prime }}J_{\eta \eta ^{\prime }}\bigl [\bigl(c_{\eta \mathbf{k}
\uparrow }^{\dag }c_{\eta ^{\prime }\mathbf{k}^{\prime } \uparrow}^{{\phantom{\dagger}}} - 
c_{\eta \mathbf{k}\downarrow }^{\dag }c_{\eta ^{\prime }
\mathbf{k}^{\prime }\downarrow }^{{\phantom{\dagger}}}\bigr)S^{z} \cr 
&& +c_{\eta \mathbf{k}\uparrow }^{\dag }c_{\eta ^{\prime }\mathbf{k}^{\prime} 
\downarrow}^{{\phantom{\dagger}}}S^{-} + c_{\eta \mathbf{k}\downarrow}^{\dag} 
c_{\eta^{\prime} \mathbf{k}^{\prime }\uparrow }^{{\phantom{\dagger}}} S^{+}\bigr]+ H_{\rm 
dir}, \label{tunneling} \\
H_{\rm dir} &=& J_{0} \sum_{\sigma} \bigl ( c_{L {\bf k} \sigma}^\dagger + c_{R {\bf k} 
\sigma}^\dagger \bigr ) \bigl ( c_{L {\bf k} \sigma}^\pdag + c_{R {\bf k} \sigma}^\pdag \bigr 
), \label{direct}
\en 
\end{subequations}
where $\theta=\omega_c t$, $c_{\eta \mathbf{k}\sigma }^{\dagger }$ ($c_{\eta \mathbf{k}\sigma 
}$) is the creation (annihilation) operator for electrons with momentum $\mathbf{k}$, 
spin-$\sigma$ and energy $\epsilon _{\eta \mathbf{k}}$ in lead $\eta$ ($=\mathrm{L,R}$), 
${\bf S}\equiv (S^x, S^y, S^z)$ are Pauli spin operators of electrons in the QD 
[$S^{\pm}\equiv S^x \pm i S^y$], and $J_{\eta \eta ^{\prime }}$ is the exchange coupling 
constant.
$\Delta_{0}=g\mu _{B}B_0$ is the static magnetic-field $B_0$-induced Zeeman term, and 
$\Delta=g\mu _{B}B_1$ describes the spin-flip scattering caused by the CPF. $H_{\rm dir}$ is 
the potential scattering term, which is decoupled from the electron spin due to the number of 
electrons in the dot level being one. As a result, this term has no influence on the 
dynamical evolution of the electron spin and behaves only as a direct bridge to connect the 
left and right leads. For an Anderson model with symmetrical coupling to the leads $t$, we 
have $J_{\eta \eta'}=2J_{0}=t^2/\epsilon_d$. As in our previous paper\cite{Dong}, we can 
rewrite the tunneling term, Eq.~(\ref{tunneling}), as a sum of three products of two 
variables: $H_{\rm I}=Q^z S^z+ Q^+ S^- + Q^- S^+ + Q^{\hat 1}$ with the same definitions of 
$Q^{z(\pm)}$ as in Ref.~\onlinecite{Dong} and $Q^{\hat 1}=H_{\rm dir}$.

It is noted that our model describes the non-resonant tunneling through a localized magnetic 
impurity involving s-d exchange interaction with the conduction electrons in the presence of 
both a static magnetic field ${\bf B}_0$ and a rotating magnetic field ${\bf B}_1$. We assume 
the exchange interaction to be weak so that no Kondo effect emerges, and assume that charge 
fluctuation completely vanishes.
 

It is well-known that the isolated spin-$1/2$ electron under the influence of both ${\bf 
B}_0$ and an external CPF ${\bf B}_1(t)$ (the Rabi problem\cite{Rabi}), Eq.~(\ref{QD}), has 
an analytical solution.\cite{Grifoni,Allen} In the rotating frame, $x=\cos \theta S^x - \sin 
\theta S^y$, $y=\sin \theta S^x + \cos \theta S^y$, and $z=S^z$, the QD Hamiltonian 
Eq.~(\ref{QD}) takes the form
\bq
\widetilde H_0=-\Delta x- \delta z,
\eq
with $\delta=\Delta_0-\omega_c$. Correspondingly, the Heisenberg equations of motion (EOM's) 
of these free rotating coordinates become: $\dot x= \delta y$, $\dot y=-\delta x + \Delta z$, 
and $\dot z=-\Delta y$. Solving these resulting ordinary differential equations, we obtain 
the free evolutions of these rotating coordinates:
\bn
{\bf r}(t') &=& {\bf M}(\tau) {\bf r}(t), \,\,\,[{\bf r}(t)\equiv 
(x(t),y(t),z(t))^T],\label{freeevolution} \\ 
{\bf M}(\tau) &=& \left ( 
\begin{array}{ccc}
\frac{\delta^2}{\Omega^2} a^+ + \frac{\Delta^2}{\Omega^2} & -i \frac{\delta}{\Omega} a^- & 
-\frac{\delta \Delta}{\Omega^2} (a^+-1) \\
i \frac{\delta}{\Omega} a^- & a^+ & -i \frac{\Delta}{\Omega} a^- \\
-\frac{\delta \Delta}{\Omega^2} (a^+-1) & i \frac{\Delta}{\Omega} a^- & 
\frac{\Delta^2}{\Omega^2} a^+ + \frac{\delta^2}{\Omega^2}
\end{array}
\right ), \nonumber
\en
with $\tau=t-t'$, $a^{\pm}=\frac{1}{2}(e^{-i \Omega \tau} \pm e^{i\Omega \tau})$, and the 
Rabi frequency $\Omega=\sqrt{\Delta^2 + \delta^2}$. 

Furthermore, the transformed interacting Hamiltonian, Eq.~(\ref{tunneling}), reads in terms 
of these rotating frame:
\bq
\widetilde H_{\rm I}= Q^x x + Q^y y + Q^z z + Q^{\hat 1}, \label{H_I}
\eq
with $Q^x=e^{-i\theta} Q^- + e^{i\theta} Q^+$ and $Q^y=i (e^{-i\theta} Q^- - e^{i\theta} 
Q^+)$. The Heisenberg EOM's for the spin operators are:
\begin{subequations}
\label{Heisenberg}
\bn
\dot x &=& \delta y - Q^z y + Q^y z, \\
\dot y &=& -\delta x + \Delta z + Q^z x - Q^x z , \\
\dot z &=& -\Delta y - Q^y x + Q^x y.
\en     
\end{subequations}
It is clear that the spin dynamics, apart from free evolutions, are perturbatively modified 
by the weak tunnel coupling. To obtain the modified dynamics, we employ a generic quantum 
Langevin equation approach.\cite{Dong,Ackerhalt,Smirnov} In our derivation, operators of the 
QD spin and the reservoirs are first expressed formally by integration of their Heisenberg 
EOM's, Eq.~(\ref{Heisenberg}), exactly to all orders of $J_{\eta\eta'}$. Next, under the 
assumption that the time scale of decay processes is much slower than that of free 
evolutions, we replace the time-dependent operators involved in the integrals of these EOM's 
approximately in terms of their free
evolutions, Eq.~(\ref{freeevolution}). Thirdly, these EOM's are expanded in powers of 
$J_{\eta\eta'}$ up to second
order. To this end, we can establish the Bloch-type dynamical equations for the averaged spin 
variables ${\bf r}$ as:
\bq
\left ( 
\begin{array}{c}
\dot x \\
\dot y \\
\dot z
\end{array}
\right ) = \left ( 
\begin{array}{ccc}
-\Gamma_{xx} & \delta & \Gamma_{xz} \\
-\delta & -\Gamma_{yy} & \Delta \\
\Gamma_{zx} & -\Delta & -\Gamma_{zz}
\end{array}
\right ) \left ( 
\begin{array}{c}
x \\
y \\
z
\end{array}
\right ) + \left ( 
\begin{array}{c}
\gamma_x \\
0 \\
\gamma_z
\end{array}
\right ), \label{Blocheq}
\eq
in which
\begin{subequations}
\bn
\Gamma_{xx} &=& 2\left ( \frac{\Delta}{\Omega}\right )^2 C(\Omega) + 2\left ( 
\frac{\delta}{\Omega}\right )^2 C(0) \cr
&& \hspace{-1.cm} + \left ( 1- \frac{\delta}{\Omega} \right ) C(\omega_c-\Omega) + \left ( 1+ 
\frac{\delta}{\Omega}\right ) C(\omega_c+\Omega), \\
\Gamma_{xz} &=& 2 \frac{\delta \Delta}{\Omega^2} [ C(0) - C(\Omega) ], \\
\gamma_x &=& \frac{\Delta}{2 \Omega} \left [ 2 R(\Omega) - 2 \frac{\delta}{\Omega} 
R(\omega_c) \right.  
- \left ( 1- \frac{\delta}{\Omega}\right ) R(\omega_c-\Omega) \cr
&& \left. + \left ( 1+ \frac{\delta}{\Omega}\right ) R(\omega_c+\Omega)\right ], \\
\Gamma_{yy} &=& 2\left ( \frac{\Delta}{\Omega}\right )^2 [C(\Omega)+ C(\omega_c)] + 2\left ( 
\frac{\delta}{\Omega}\right )^2 C(0)  \cr
&& \hspace{-1cm} + \frac{\delta}{\Omega} \left [ \left (1+ \frac{\delta}{\Omega} \right ) 
C(\omega_c+\Omega) - \left ( 1- \frac{\delta}{\Omega}\right ) C(\omega_c-\Omega) \right ], 
\cr
&& \\
\Gamma_{zx} &=& 2\frac{\delta \Delta}{\Omega^2} C(\omega_c) + \frac{\Delta}{\Omega} \left [ 
\left ( 1- \frac{\delta}{\Omega}\right ) C(\omega_c-\Omega) \right. \cr 
&& \left. - \left (1+ \frac{\delta}{\Omega}\right ) C(\omega_c+\Omega) \right ],\\
\Gamma_{zz} &=& 2\left ( \frac{\Delta}{\Omega}\right )^2 C(\omega_c) + \left (1- 
\frac{\delta}{\Omega}\right )^2 C(\omega_c-\Omega) \cr
&& + \left (1+ \frac{\delta}{\Omega}\right )^2 C(\omega_c+\Omega),\\
\gamma_z &=& \frac{1}{2} \left [ 2 \left ( \frac{\Delta}{\Omega}\right )^2 R(\omega_c) + 
\left ( 1- \frac{\delta}{\Omega}\right )^2 R(\omega_c-\Omega) \right. \cr
&& \left. + \left (1+ \frac{\delta}{\Omega}\right )^2 R(\omega_c+\Omega)\right ], 
\en
\end{subequations}
with the reservoir correlation function, $C(\omega)$, and the response function, $R(\omega)$, 
defined as
\begin{subequations}
\bn
C(\omega)&=& \frac{\pi}{2} \left ( g_{LL}+ g_{RR} \right ) T \varphi \left ( \frac{\omega}{T} 
\right ) \cr
&& \hspace{-1.5cm}+ \frac{\pi}{2} g_{LR} T \left [ \varphi \left ( \frac{\omega + V}{T}\right 
) + \varphi \left ( \frac{\omega- V}{T}\right ) \right ], \label{C}
\\
R(\omega)&=& \frac{\pi}{2} \left ( g_{LL}+ g_{RR} + 2g_{LR} \right ) \omega, 
\en
\end{subequations}
with $g_{\eta\eta'}\equiv J_{\eta\eta'}^2 \rho_0^2$ ($\rho_0$ is the constant density of 
states of both electrodes) and $\varphi(x)\equiv x \coth (x/2)$. $V$ is the bias-voltage 
applied symmetrically to the two electrodes, $\mu_L=-\mu_R=V/2$. We use units with 
$\hbar=k_B=e=1$. 

Notice that these quantities of $\Gamma_{\alpha\beta}$ describe the dissipation of 
cotunneling processes to the dynamics of the QD spin, in conjunction with the effect of CPF: 
transition between two spin-splitting states $\pm \Omega/2$ (spin analog of charge optical 
Stark effect) due to photo-absorption and/or emission. Also, it is easy to see that the Bloch 
equations, Eq.~(\ref{Blocheq}), is exactly reduced to the results of Ref.~\onlinecite{Dong} 
in the case of vanishing CPF, $\Delta=0$ and $\omega_c=0$.       

The nonequilibrium steady-state spin projections of the QD in the rotating frame, ${\bf 
r}^{\infty} = (x^{\infty}, y^{\infty}, z^{\infty})$, can be readily obtained from 
Eq.~(\ref{Blocheq}). As is evident, $x^\infty=y^\infty=0$, while 
$z^\infty=\frac{1}{2}R(\Delta_0)/C(\Delta_0)$ is identical to the previous theoretical result 
of nonequilibrium magnetization\cite{Dong,Parcollet} in absence of driving CPF. On the other 
hand, in the case of vanishing static magnetic field, $\Delta_0=0$, the nonzero driving CPF 
can induce an additional spin orientation $z^\infty\neq 0$ along the rotating direction of 
the CPF, together with the nonzero $x^\infty$ (not shown here). At equilibrium, this 
optically-induced spin orientation phenomenon has historically been termed, in literature, as 
either the inverse Faraday effect\cite{acStark1} or as the Zeeman light shift,\cite{acStark2} 
which is ascribed to the CPF-induced spin-splitting (ac spin Stark effect). In addition, one 
can interestingly observe a nearly vanishing $y$-component of the spin polarization even at 
CPF driving.

\begin{figure*}[htb]
\begin{center}
\includegraphics[width=17cm,height=8cm,angle=0,clip=on]{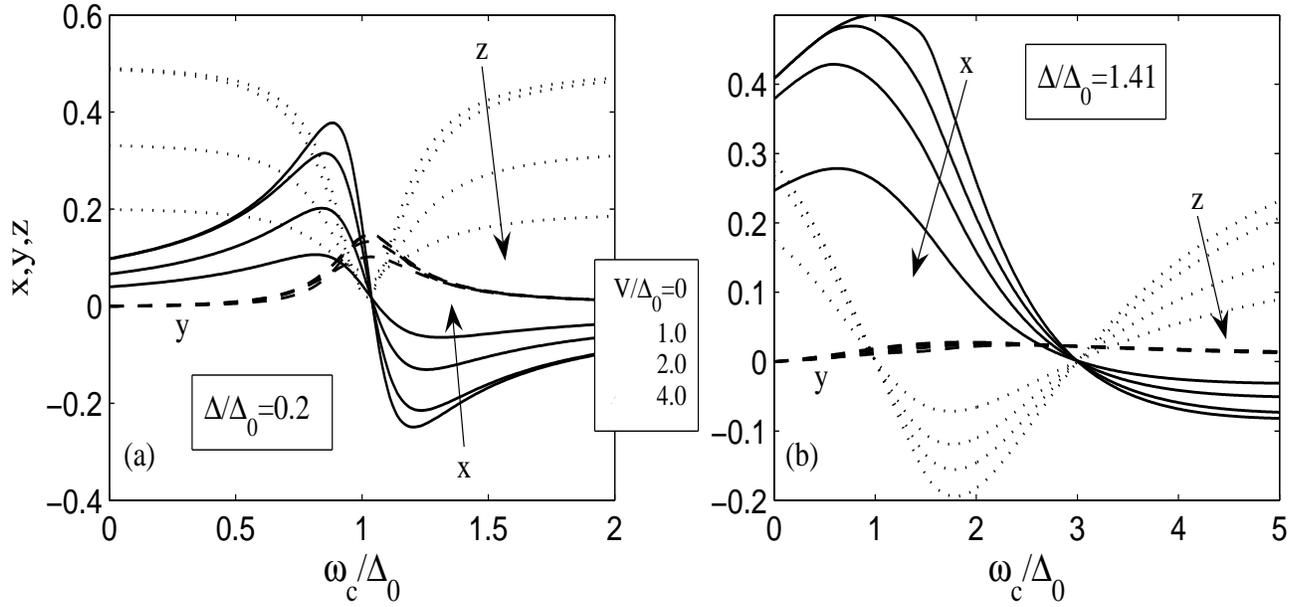}                         
\caption{The nonequilibrium spin projection ${\bf r}^\infty$ vs. $\omega_c/\Delta_0$ 
($\Delta_0=1$) with increasing bias-voltage under a weak driving CPF, $\Delta=0.2 \Delta_0$ 
(a), and a strong driving field, $\Delta=\sqrt{2}\Delta_0$ (b). The arrow indicates the 
direction of increase of bias-voltage $V$. The other parameters are: 
$g_{LL}=g_{RR}=g_{LR}=0.05$ and $T=0.02$.} \label{soo}
\end{center}
\end{figure*}

For nonzero static magnetic field, previous theoretical studies show that coherent 
suppression of tunneling may take place only if the Rabi frequency $\Omega$ matches with the 
driving frequency $\omega_c$, which corresponds to the following relationship of the driving 
field\cite{Shao}:
\bq
\Omega^*=\omega_c=\frac{\Delta_0^2+ \Delta^2}{2\Delta_0}. \label{localization}
\eq
In Fig.~1, we plot the nonequilibrium spin projection ${\bf r}^\infty$ under the influence of 
both nonzero static magnetic field and driving CPFs. Different from the results of 
$\Delta_0=0$, we find an obvious nonzero polarization of the $y$-component of QD spin near 
the resonant field, $\omega_c= \Delta_0$, for the weak driving field $\Delta=0.2\Delta_0$ 
[Fig.~1(a)]. Interestingly, we observe $z^\infty=0$ at $\omega_c=\Delta_0$ for any strength 
of CPF, meaning that the resonant frequency of a CPF may overcome the Zeeman-splitting of a 
static magnetic field. At the same time, $x^\infty$ becomes unpolarized for weak driving 
field [Fig.~1(a)], while reaches the maximum value for strong driving field, which decreases 
with increasing of the bias-voltage [Fig.~1(b)]. Moreover, we notice that (1) 
$x^\infty=z^\infty=0$ at $\omega_c=2\Omega^*$; and (2) $z^\infty$ is negative if 
$\Delta_0<\omega_c<2\Omega^*$ for strong driving CPF.

We now proceed with the calculation of tunneling current. The current operator through the QD 
is defined as the time rate of change of charge density $N_{\eta}=\sum_{{\bf k},\sigma} 
c_{\eta{\bf k}\sigma}^\dagger c_{\eta{\bf k} \sigma}^\pdag$ in lead $\eta$ (we choose the 
left lead as an example): 
\bn
J_{L }(t) &=& -\dot N_{L }=i[N_{L},H]_- \cr 
&=& i(Q_{LR}^{z\uparrow\uparrow}-Q_{RL}^{z\uparrow\uparrow})S^z - 
i(Q_{LR}^{z\downarrow\downarrow}-Q_{RL}^{z\downarrow\downarrow})S^z \cr
&& +i(Q_{LR}^- - Q_{RL}^-)S^+ +i(Q_{LR}^+ - Q_{RL}^+)S^- \cr
&& + i(Q_{LR}^{z\uparrow\uparrow}-Q_{RL}^{z\uparrow\uparrow} + 
Q_{LR}^{z\downarrow\downarrow}-Q_{RL}^{z\downarrow\downarrow}),  \label{current-operator}
\en
where the definitions of $Q_{\eta\eta'}^{z\sigma\sigma}$ and $Q_{\eta\eta'}^{\pm}$ can be 
found in our previous paper\cite{Dong} and those terms in the last line are stemming from the 
direct tunneling term $H_{\rm dir}=J_0 \sum_{\eta\eta',\sigma} 
Q_{\eta\eta'}^{z\sigma\sigma}$. 
The linear-response theory gives
\bq
I = \langle J_{L }(t) \rangle =-i\int_{-\infty}^t dt' \langle [ J_{L }(t), H_{\rm I}(t')]_- 
\rangle_{0},  \label{current}
\eq
where the statistical average $\langle \cdots \rangle_0$ is performed with respect to 
decoupled two subsystems, QD and reservoirs. Inserting Eqs.~(\ref{tunneling}) and 
(\ref{current-operator}) into Eq.~(\ref{current}), one can derive the explicit expression for 
steady-state current in terms of the steady-state spin projections ${\bf r}^{\infty}$ through 
a lengthy but straightforward calculation. In particular, we note
\bn
\langle [(Q_{LR}^{z\uparrow\uparrow} -Q_{LR}^{z\downarrow\downarrow})S^z, H_{\rm dir}]_- 
\rangle_0 &=& 0, \\
\langle [(Q_{RL}^{z\uparrow\uparrow} -Q_{RL}^{z\downarrow\downarrow})S^z, H_{\rm dir}]_- 
\rangle_0 &=& 0, \\
\langle [Q_{\eta\eta'}^{\pm} S^{\mp}, H_{\rm dir}]_- \rangle_0 &=& 0,
\en
and
\bn
&& \langle [(Q_{LR}^{z\uparrow\uparrow}-Q_{RL}^{z\uparrow\uparrow} + 
Q_{LR}^{z\downarrow\downarrow}-Q_{RL}^{z\downarrow\downarrow}), H_{\rm I}] \rangle_0\cr
&& = \langle [(Q_{LR}^{z\uparrow\uparrow}-Q_{RL}^{z\uparrow\uparrow} + 
Q_{LR}^{z\downarrow\downarrow}-Q_{RL}^{z\downarrow\downarrow}), H_{\rm dir}] \rangle_0 \neq 
0,
\en
because the QD is connected to two normal leads in the system under consideration.\cite{Dong}
These results indicate that the contribution of the direct tunneling term to the current is 
independent of the dynamics of the QD and only depends on the bias-voltage and temperature of 
the two electrodes.
To this end, the current $I$ is\cite{DongE}
\begin{widetext}
\bn
\frac{I}{\pi g_{LR}}&=& 4 V - T \left \{ \frac{1}{2} \left [ \left ( 
\frac{\delta}{\Omega}\right )^2 +1 \right ] {\cal I}^+ + \frac{\delta}{\Omega} {\cal I}^- + 
\left ( \frac{\Delta}{\Omega}\right )^2 \left [ \varphi \left ( \frac{\omega_c+V}{T} \right ) 
- \varphi \left ( \frac{\omega_c-V}{T} \right ) \right ] \right \} z^{\infty} \cr
&& \hspace{-1cm} - 2T \left \{ \frac{\delta\Delta}{4\Omega^2} {\cal I}_c^+ + 
\frac{\Delta}{4\Omega} {\cal I}_c^- - \frac{\delta\Delta}{2\Omega^2} \left [ \varphi \left ( 
\frac{\omega_c+V}{T} \right ) - \varphi \left ( \frac{\omega_c-V}{T} \right ) \right ] + 
\frac{\Delta}{2\Omega} \left [ \varphi \left ( \frac{\Omega+V}{T} \right ) - \varphi \left ( 
\frac{\Omega-V}{T} \right ) \right ] \right \} x^{\infty}, \label{ic}
\en
\end{widetext}
with
\bn
{\cal I}^{\pm} &=& \varphi \left ( \frac{\omega_c+\Omega+V}{T} \right ) \pm \varphi \left ( 
\frac{\omega_c -\Omega+V}{T} \right ) \cr
&& \hspace{-1cm} - \left [ \varphi \left ( \frac{\omega_c+\Omega-V}{T} \right ) \pm \varphi 
\left ( \frac{\omega_c-\Omega-V}{T} \right ) \right ]. \label{calI}
\en
Different from the previous results of the nondriving QD system,\cite{Dong} the nonzero 
$x$-component of the stationary spin polarization has additional contribution to the current 
besides the $z$-component. The linearly bias-voltage-dependent term in Eq.~(\ref{ic}) stems 
from the cotunneling processes, in which the spin projection remain unchanged. The CPF has no 
effect on these processes.
More importantly, one observes that the current formula Eq.~(\ref{ic}) includes some new 
factors involving $\omega_c\pm \Omega \pm V$ [Eq.~(\ref{calI})], which can be intuitively 
ascribed to the {\em photo-assisted} spin-flip cotunneling processes. For instance, the term 
involving $\Omega-\omega_c$ results from the contribution of cotunneling accompanied by an 
{\em absorption} of one-photon [Fig.~2(a)], while the term involving $\Omega+\omega_c$ 
corresponds to the process assisted by a spontaneous {\em emission} of one-photon 
[Fig.~2(b)]. Therefore, we simply name these cotunnelings as $\Omega\pm \omega_c$ processes 
respectively below. Besides, the factors $\Omega\pm V$ in Eq.~(\ref{ic}) stem from the 
spin-flip cotunneling events without spontaneous emission or absorption of photon, while the 
role of driving field is reflected through the effective spin-splitting $\Omega$. In the 
following, we will see that the joint effects of these new terms are responsible for the 
controllable patterns of cotunneling through a QD by the presence of CPF.           

\begin{figure}[htb]
\begin{center}
\includegraphics[width=8cm,height=4cm,angle=0,clip=on]{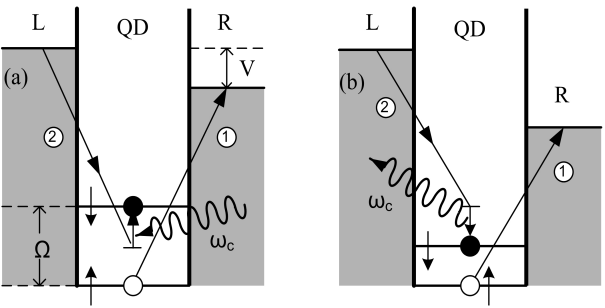}
\caption{Schematic diagrams of photon-absorption- ($\Omega-\omega_c$) (a) and 
photon-emission- ($\Omega+\omega_c$) (b) assisted spin-flip cotunneling processes.} 
\label{fig2}
\end{center}
\end{figure}

\begin{figure*}[htb]
\begin{center}
\includegraphics[width=17cm,height=12cm,angle=0,clip=on]{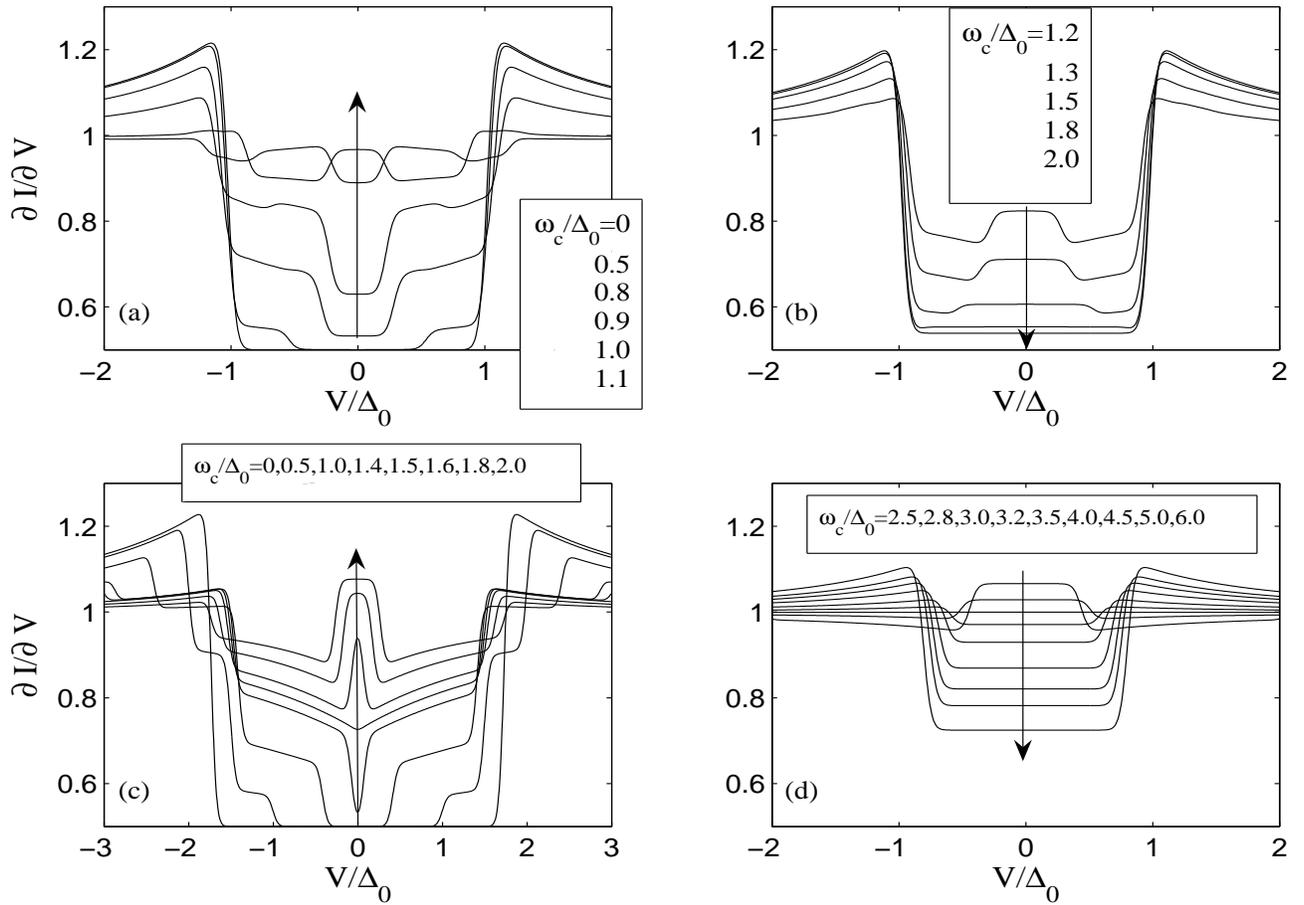}                                             
\caption{The calculated differential conductance $dI/dV$ vs. bias-voltage $V/\Delta_0$ 
($\Delta_0=1.0$) under weak driving CPF $\Delta/\Delta_0=0.2$ (a,b) and strong CPF 
$\Delta/\Delta_0=\sqrt{2}$ (c,d) with several driven frequencies at a nonzero static magnetic 
field. The arrow indicates the direction of increase of driven frequency $\omega_c$. The 
other parameters are the same as in Fig.~1.} \label{figic}
\end{center}
\end{figure*}

We plot the differential conductance $G=\frac{\partial I}{\partial V}$ (in unit of $2\pi 
g_{LR}$) vs bias-voltage as functions of different driving frequencies $\omega_c$ for a weak 
driving CPF $\Delta/\Delta_0=0.2$ (a,b) and a strong driving field $\Delta/\Delta_0=\sqrt{2}$ 
(c,d) in Fig.~\ref{figic}. If $\omega_c=0$, the $I$-$V$ curve reduces to the ordinary 
cotunneling characteristics under an ambient constant magnetic field:\cite{Dong} a jump at 
$V=\pm \Omega = \pm \sqrt{\Delta_0^2+ \Delta^2}$ [$\approx \Delta_0$ in Fig.~\ref{figic}(a) 
due to $\Delta \ll \Delta_0$ and $=\sqrt{3} \Delta_0$ in Fig.~\ref{figic}(c), respectively], 
ascribing to the energetic inactivation (activation) of spin-flip processes at bias-voltage, 
$|V|< (>)\Omega$. While for nonzero driving frequency, such as $\omega_c/\Delta_0=0.5$, the 
CPF effectively suppresses spin-splitting $\Omega/\Delta_0 \approx 0.54$ at the case of weak 
driving field, thus leading to two jumps, one of which is located at $V=\pm \Omega$ due to 
the pure activated spin-flip process, and another of which appears at $V=\pm 
(\Omega+\omega_c)\approx \pm \Delta_0$, which can be ascribed to opening of an additional 
channel for electron transfer cotunneling due to the one-photon-emission-assisted spin-flip 
resonant event. In contrast, the strong driving field $\Delta=\sqrt{2} \Delta_0$ makes 
$\Omega=1.5 \Delta_0$ and generates three jumps: the first one corresponds to the 
$\Omega-\omega_c$ process; the second one is due to the pure electronic spin-flip event 
($\Omega$); and the third one results from the $\Omega+\omega_c$ process.  
When the frequency increases to a bit higher than the resonant frequency [$1.0 \Delta_0$ 
($\Omega^*=1.5\Delta_0$) for the weak (strong) driving field], a zero-bias peak emerges for 
both cases, which can be understood by the fact that the proper high frequency CPF has enough 
energy to spur the spin-flip cotunneling even at equilibrium [see Fig.~2(a)], while the 
rising voltage, $V\geq \pm \Omega$ [Figs.~\ref{figic}(a, b)] or $\pm (\Omega-\omega_c)$ 
[Figs.~\ref{figic}(c, d)], contrarily suppresses its activation until a pure bias-driven 
spin-flip process is excited at $V=\pm \Delta_0$ or $\pm \Omega$. Recently, the zero-anomaly 
(ZA) behavior of cotunneling current has been also reported for a QD connected to two 
anti-parallel ferromagnetic electrodes.\cite{Weymann,Dong2} Here, it  should be pointed out 
that the appearance of ZA in the present system is due to photon-assisted {\em inelastic} 
spin-flip scattering becoming resonant in the presence of external magnetic field, which is 
different from the previous mechanism, {\em elastic} spin-flip event in the case of 
polarization leads without any magnetic field.\cite{Weymann,Dong2} As a result, the present 
system has more rich transport features by tuning strength and frequency of driving CPF. 

For the weak CPF field with $\omega_c \approx 2\Delta_0$, the photon-assisted excitation of 
spin-flip cotunneling merges coincidently into pure bias excitation, leading to the 
disappearance of ZA behavior [Fig.~\ref{figic}(b)]. While for strong CPF field 
[Fig.~\ref{figic}(d)], the ZA also vanishes if $\omega_c \geq 2\Omega^*$, which is due to the 
peculiar features of the nonequilibrium spin projections $x^\infty$ and $z^\infty$ [see 
Fig.~1(b)]. For higher driving frequency, the cotunneling exhibits splitting differential 
conductance and eventually tends to the pattern in the presence of a static magnetic field 
alone, because the QD spin cannot, from physical point of view, catch the details of the 
driving field with considerably high frequency. Finally, we point out that the ZA feature is 
robust over a wide region of temperature (not shown here).                          

\textit{Conclusion.}---In summary, we have presented an analytical study of the inelastic 
cotunneling, including the nonequilibrium spin projections and currents, in a single QD under 
a static magnetic field and a perpendicular CPF, revealing a controllable $I$-$V$ pattern, 
the transition between ZA and splitting of the differential conductance, due to {\em 
photon-assisted spin-flip} inelastic cotunneling.

This work was supported by Projects of the National Science Foundation of China and
the Shanghai Municipal Commission of Science and Technology, the Shanghai Pujiang Program, 
and NCET.


\begin{thebibliography}{99}
\bibitem{DongE}{We took no account of the contribution of the direct tunneling term in our 
previous paper, Ref.~\onlinecite{Dong}. If this term is included, the linearly bias-voltage 
relevant term in the charge current formula, Eq.~(36), becomes $4V$ in stead of $3V$.}
\end{thebibliography}
\end{document}